\pdfoutput=1

\documentclass[11pt]{article}

\usepackage[final]{acl}

\usepackage{times}
\usepackage{latexsym}

\usepackage[T1]{fontenc}

\usepackage[utf8]{inputenc}

\usepackage{microtype}

\usepackage{inconsolata}

\usepackage{graphicx}

\usepackage{amsmath} 
\usepackage{booktabs}
\newcommand{\lqy}[1]{{#1}}

\title{Grammar-Based Code Representation: Is It a Worthy Pursuit for LLMs?}

\author{
 \textbf{Qingyuan Liang\textsuperscript{1,2}$^\dagger$},
 \textbf{Zhao Zhang\textsuperscript{1,2}$^\dagger$},
 \textbf{Zeyu Sun\textsuperscript{3*}},
 \textbf{Zheng Lin\textsuperscript{2}},
 \textbf{Qi Luo\textsuperscript{4}},
 \textbf{Yueyi Xiao\textsuperscript{1}},\\
 \textbf{Yizhou Chen\textsuperscript{1}},
 \textbf{Yuqun Zhang\textsuperscript{4}},
 \textbf{Haotian Zhang\textsuperscript{2}},
 \textbf{Lu Zhang\textsuperscript{1}},
 \textbf{Bin Chen\textsuperscript{2}},
 \textbf{Yingfei Xiong\textsuperscript{1*}}
\\
\\
 \textsuperscript{1}School of Computer Science, Peking University,
 \textsuperscript{2}Kuaishou Technology,\\
 \textsuperscript{3}Institute of Software, Chinese Academy of Sciences,
 \textsuperscript{4}Department of Computer Science \\and Engineering, Southern University of Science and Technology
 \\
 \small{
 \{liangqy, zhangzhao2019, xiaoyueyi, yizhouchen, zhanglucs, xiongyf\}@pku.edu.cn; zeyu.zys@gmail.com};
 \\
 \small{ \{linzheng, zhanghaotian, chenbin\}@kuaishou.com};
 \small{\{12232440, zhangyq\}@sustech.edu.cn}
 \\
}

\renewcommand{\thefootnote}{}

\begin{document}
\maketitle

\footnotetext{$^\dagger$ Work done during internship at Kuaishou Technology.}
\footnotetext{*Corresponding authors: Zeyu Sun and Yingfei Xiong.}
\renewcommand{\thefootnote}{\arabic{footnote}}

\begin{abstract}
Grammar serves as a cornerstone in programming languages and software engineering, providing frameworks to define the syntactic space and program structure.
Existing research demonstrates the effectiveness of grammar-based code representations in small-scale models, showing their ability to reduce syntax errors and enhance performance.
However, as language models scale to the billion level or beyond, syntax-level errors become rare, making it unclear whether grammar information still provides performance benefits.
To explore this, we develop a series of billion-scale GrammarCoder models, incorporating grammar rules in the code generation process. 
Experiments on HumanEval~(+) and MBPP~(+) demonstrate a notable improvement in code generation accuracy. Further analysis shows that grammar-based representations enhance LLMs' ability to discern subtle code differences, reducing semantic errors caused by minor variations. 
These findings suggest that grammar-based code representations remain valuable even in billion-scale models, not only by maintaining syntax correctness but also by improving semantic differentiation.
\end{abstract}

\begin{figure*}[t]
  \includegraphics[width=\linewidth]{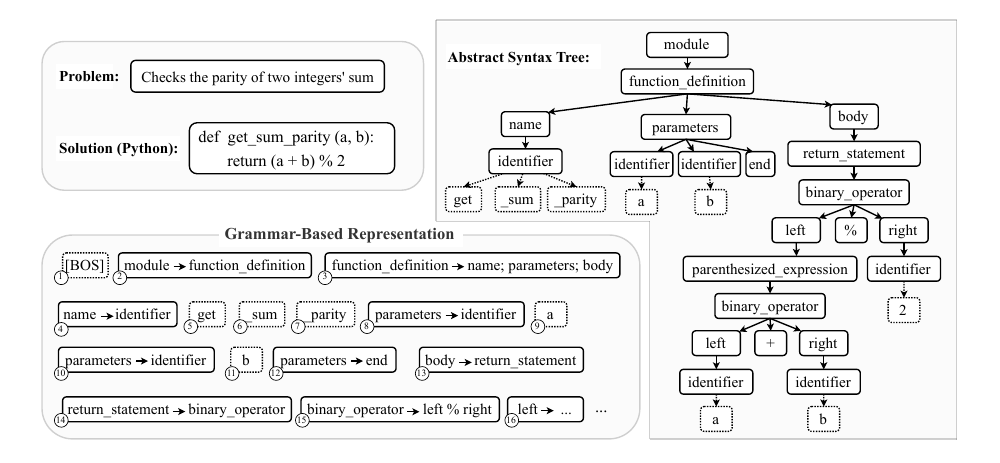} 
  \caption {An example of a grammar representation. The top-left part presents a programming problem along with its corresponding Python solution. The right part illustrates the abstract syntax tree~(AST) representation of the Python code. The bottom-left section presents the grammar-based representation.}
  \label{grammar}
  \vspace{-0.3cm}
\end{figure*}

\section{Introduction}

Context-free grammars are the fundamental way to specify the syntactic space of a programming language, and with the grammar specified, a program can be parsed into a syntax tree, revealing its structure~\cite{compiler}. Building on this foundation, leveraging grammatical knowledge~(e.g., grammar rules) to pre-train large language models~(LLMs) has emerged as a promising strategy for code-related tasks, such as code generation~\cite{grammart5, treegen, graphcodebert2020}.

Existing research has explored grammar-based code representation~\cite{treebert,unixcoder,syncobert, grammart5, treegen, l2s, abstract}, where each grammar rule serves as an identity token, and a sequence of grammar rules and terminal tokens represents the program.
Figure~\ref{grammar} illustrates a program that determines whether the sum of two integers is odd~(top left), along with its corresponding abstract syntax tree~(AST) representation~(right) and grammar-based representation~(bottom left). The grammar-based representation is derived by performing a preorder traversal on the AST. Each grammar rule is extracted independently~(e.g., `module $\rightarrow$ function\_definition'), while terminals are tokenized using a standard tokenizer~(e.g., `get').
Grammar-based representation has been shown to be effective in preventing syntax errors in encoder-decoder architecture~\cite{grammart5}. Moreover, it facilitates program analysis and enables the pruning of incorrect branches (e.g., filtering out type-error programs~\cite{l2s,tare}) during code generation, thereby enhancing accuracy.
Due to these benefits, many code generation models adopt grammar-based representation~\cite{sun2019grammar, grammart5}.


However, as language models scale to the billion-parameter level and beyond, extensive pre-training on large code datasets enables them to implicitly learn syntax rules, making syntax errors increasingly rare~\cite{chatgpt2024, qwen2, qwen2.5, deepseekcoderv2}.
For example, even 1B scale models, such as DeepSeek-Coder~\cite{guo2024deepseekcoder} and Qwen2.5~\cite{qwen2.5}, achieve high accuracy in code generation, consistently producing syntactically valid code. 
This phenomenon suggests that large models are able to understand the structure of the program and raises a critical question: \textit{Is grammar-based code representation still beneficial for billion-scale LLMs?}

To answer this question, we conduct an experiment comparing grammar-based representation and token-based representation approaches on 1.3B, 1.5B, and 7B parameter models, respectively. 
The results demonstrate that grammar-based models~(i.e., GrammarCoder-1.3B-Base, GrammarCoder-1.5B-Base, GrammarCoder-7B-Base) significantly outperform token-based models, even though token-based models rarely make syntax errors. 
For example, on the MBPP dataset~\cite{mbpp}, GrammarCoder-1.3B-Base achieves an almost seven percentage point improvement in Pass@1 compared to DeepSeek-Coder-1.3B-Base trained on the same data.
This suggests that grammar rules enhance code generation beyond syntax correction, even in billion-scale models.

The result leads to a second question: \textit{Why do grammar-based models improve performance if token-based models already produce syntactically correct code?} 


To investigate this question, we examine the differences between grammar-based and token-based code representations. Our analysis reveals that minor token-level modifications can lead to substantial semantic shifts, rendering correct programs incorrect. In contrast, while these subtle variations may appear insignificant at the token level, they often map to clear structural differences in grammar-based representations, enabling the model to distinguish more effectively between correct and incorrect code. Experimental results further confirm a correlation between higher performance and 
the ability of grammar-based code representation to amplify representational differences for semantic shifts, indicating that grammar-based representation helps mitigate such semantic issues.

Our main contributions are as follows:
\begin{itemize}
    \item We are the first to conduct an experiment on grammar-based code representation in billion-scale LLMs, finding that it remains effective compared to token-based approaches.
    \item We are the first to explain the effectiveness of grammar-based representation beyond syntax correctness and validate our hypothesis through empirical experiments, demonstrating its role in enhancing code semantic differentiation.
    \item We release a series of code LLMs trained with grammar rules, providing a valuable resource for further research~\cite{grammarcoder}.
\end{itemize}

\section{GrammarCoder}
\subsection{Model Overview}

We propose GrammarCoder, a grammar-based model built on a decoder-only architecture~\cite{transformer2017, gpt2018}, which excels in auto-regressive tasks like code generation, completion, and translation~\cite{chatgpt2024, guo2024deepseekcoder, qwen2.5, qwen2.5coder, deepseekcoderv2}. 
To enhance its ability to code generation, we apply continued pre-training and instruction tuning on existing code model weights~(i.e., DeepSeek-Coder-1.3B-Base, Qwen2.5-1.5B-Base, and Qwen2.5-7B-Base), expanding its knowledge base.
\ref{mode_config} provides the configuration of the base model we used.
In this section, we first introduce our grammar-based code representation. Then, we describe our training strategy and corpus.

\subsection{Grammar-Based Code Representation}
The main idea of grammar-based code representation is to guide the model in generating grammar rules rather than merely producing a sequence of normal tokens. Traditional code LLMs primarily rely on token-level composition to construct complete code text. In contrast, grammar-based models first generate a complete AST by composing grammar rules and then reconstruct executable code from it, thereby enhancing the model’s understanding of code structure and logic.
\lqy{
Specifically, normal tokens are obtained using the byte pair encoding (BPE) algorithm, which learns tokenized representations from text corpora, forming a vocabulary \( V_{\text{normal}} = \{t_1, t_2, \dots, t_m\} \). This follows the standard approach used in natural language model training.
To integrate grammar information in code representation, we introduce grammar rule sequences, which represent the step-by-step derivation process of an AST. We define a grammar rule vocabulary \( V_{\text{rule}} = \{r_1, r_2, \dots, r_k\} \), where each rule encodes a structural transformation in code generation. Unlike token-based representations, grammar rule sequences explicitly capture logical dependencies and hierarchical structures, providing a more structured view of code.  
}
By integrating normal tokens~\( V_{\text{normal}} \) with grammar rules \( V_{\text{rule}} \), the model can leverage syntactic rules to strengthen its understanding of code structure.
For example, in the bottom-right section of Figure~\ref{grammar}, the solid-boxed elements represent grammar rules that guide the construction of the AST~(e.g., `parameters $\rightarrow$ identifier'), ensuring that the generated structure adheres to syntax constraints. Meanwhile, the dashed-boxed elements denote normal tokens~(e.g., `get' and `a'), which fill in leaf nodes such as variable names and string literals. These tokens can be directly reused from existing BPE tokenization, preserving syntactic correctness while maintaining flexibility in code generation.

GrammarCoder assigns a unique ID to each normal token and grammar rule, storing them in one vocabulary. 
For example, in the first 10 tokens of Figure~\ref{grammar}, IDs 2, 3, 4, 8, and 10 represent grammar rules, while IDs 1, 5, 6, 7, and 9 correspond to normal tokens.
Given a base model vocabulary of size \( m \) and \( k \) grammar rules, the extended vocabulary of GrammarCoder, denoted as \( V_{\text{grammar}} \), has a total size of \( m + k \). 
With this grammar-augmented vocabulary, raw code text is converted into a grammar-based representation, enabling the model to learn beyond token-level generation through syntax-aware parsing. Unlike traditional models that rely solely on normal tokens, imposing weak constraints, GrammarCoder incorporates grammar rules, aligning serialized code directly with the preorder traversal of its AST.

\subsection{Training Strategy}
We train the grammar-based code representation using a next-token prediction strategy, a fundamental approach for auto-regressive language models. The core idea is to predict the most probable next token given a prefix sequence, continuing the process until the full content is generated. 
In the training process, we treat the grammar-based representation of each code file as the training sample, using the sequence encoded by~\( V_{\text{grammar}} \). The model learns to predict the next most probable token~(whether a normal token or a grammar rule) based on the tokens generated so far. Formal descriptions in~\ref{ex_training_objective}.

This training strategy enables the model to dynamically incorporate grammar rules during code generation, allowing the final output adhere to syntax constraints and AST structures.

\subsection{Training Corpus}\label{training_dataset}
We organize our training corpus in two stages: continued pre-training and instruction tuning. 
Python is selected as the primary programming language for data collection due to its rich syntax and widespread use in diverse programming paradigms. This makes it an ideal candidate for evaluating the effectiveness of grammar-based representations. 

For continued pre-training, we sample 10B tokens of Python code from TheStackV2~\cite{starcoder2} dataset as the primary training data. Additionally, inspired by previous studies~\cite{opencoder}, we sample 0.5B tokens of self-contained code textbooks from open-source datasets~\cite{opencoder, nakamura2025aurora} to enhance the model’s adaptability to real-world interactive scenarios, bridging the gap between standard pre-training and practical applications. 

For instruction tuning, we leverage publicly available instruction datasets~\cite{opencoder, nakamura2025aurora} and employ the data synthesis~\cite{wei2024selfcodealign, wei2024magicoder} approach to collect a total of 6B tokens of instruction data. This ensures the model is better aligned with instruction-following tasks, improving its ability to handle real-world programming scenarios.  \ref{dataset_filter} provides detailed information about the training datasets.

\begin{table}[t]
  \centering
  \scalebox{0.71}{
\begin{tabular}{lcc}
\toprule
\textbf{Model}                      & \textbf{HumanEval(+)} & \textbf{MBPP(+)}  \\ \midrule
\textbf{Original}      &                &        
\\ \midrule
DeepSeek-Coder-1.3B-Base            & 34.8 (28.7)               & 56.7 (47.9)                         \\
Qwen2.5-1.5B-Base                & 37.2 (32.9)               & 60.2 (49.6)                      \\
Qwen2.5-7B-Base                & 57.9 (50.6)               & 74.9 (62.9)                      \\
\midrule
\textbf{Normal Token-Based CPT}      &                &    
\\ \midrule
DeepSeek-Coder-1.3B-Base (CPT)      & 43.9 (39.6)              & 61.4 (51.3)                    \\
Qwen2.5-1.5B-Base (CPT)             & 50.6 (42.7)              & 60.3 (51.1)                 \\ 
Qwen2.5-7B-Base (CPT)             & 68.9 (65.2)               & 81.5 (69.8)                 \\
\midrule
\textbf{Grammar-Based CPT}      &                &       
\\ \midrule
\textbf{GrammarCoder-1.3B-Base}     & 63.4 (57.3)     & 68.3 (56.9)               \\
\textbf{GrammarCoder-1.5B-Base}     & 65.9 (57.3)      & 64.8 (55.3)              \\ 
\textbf{GrammarCoder-7B-Base}     & \textbf{76.8} \textbf{(71.3)}      & \textbf{85.2} \textbf{(71.7)}               \\ 
\bottomrule
\end{tabular}
}
  \caption{\label{mainexps}
  Comparison of code generation performance between token-based and grammar-based models. The CPT refers to continued pre-training.
  }
\vspace{-0.5cm}
\end{table}

\section{Experiments}
To evaluate the performance of grammar-based code representations, we develop two sets of models, one with grammar-based code representation and one with token-based code representation. These models are built through continued pre-training from open-source code models, DeepSeek-Coder-1.3B-Base, Qwen2.5-1.5B, and Qwen2.5-7B, on high-quality code data. 
We begin by evaluating these models on code generation tasks, which are among the most widely recognized and commonly used benchmarks for assessing code-related capabilities (i.e., Experiment I in \ref{ex1}).

To further explore the differences between grammar-based and token-based representations, we analyze the reasons contributing to the performance gains of grammar-based representation~(i.e., Experiment II in \ref{ex2}).

\begin{table*}[t]
  \centering
  \scalebox{
  0.75
  }{
\begin{tabular}{lcccc}
\toprule
\textbf{Model} & \textbf{HumanEval} & \textbf{HumanEval+} & \textbf{MBPP} & \textbf{MBPP+} \\ 
\midrule
\multicolumn{5}{c}{\textbf{Base Models}}  \\
\midrule
DeepSeek-Coder-1.3B-Base~\cite{guo2024deepseekcoder}     & 34.8      & 28.7       & 56.7 & 47.9  \\
Qwen2.5-1.5B-Base~\cite{qwen2.5}     & 37.2      & 32.9       & 60.2 & 49.6  \\
OpenCoder-1.5B-Base~\cite{opencoder} & 54.3 & 49.4 & 70.6 & 58.7 \\
Yi-Coder-1.5B~\cite{yicoder} & 41.5 & 32.9 & 27.0 & 22.2 \\
CodeGemma-2B-Base~\cite{codegemmateam2024} & 26.8 & 20.7 & 55.6 & 46.6 \\
StarCoder2-3B~\cite{starcoder2} & 31.7 & 27.4 & 60.2 & 49.1 \\

CodeGemma-7B-Base~\cite{codegemmateam2024} & 44.5 & 41.5 & 65.1 & 52.4 \\
StarCoder2-7B~\cite{starcoder2} & 35.4 & 29.9 & 54.4 & 45.6 \\
Qwen2.5-7B-Base~\cite{qwen2.5}     & 57.9      & 50.6       & 74.9 & 62.9  \\
OpenCoder-8B~\cite{opencoder} & 66.5 & 63.4 & 79.9 & 70.4 \\
\midrule
\textbf{GrammarCoder-1.3B-Base }& 63.4      & 57.3       & 68.3 & 56.9  \\
\textbf{GrammarCoder-1.5B-Base} & 63.4      & 59.1       & 64.8 & 55.3  \\
\textbf{GrammarCoder-7B-Base} & \textbf{76.8}      & \textbf{71.3}       & \textbf{85.2} & \textbf{71.7}  \\

\midrule
\multicolumn{5}{c}{\textbf{Instruct Models}}   \\
\midrule

DeepSeek-Coder-1.3B-Instruct~\cite{guo2024deepseekcoder} & 65.9      & 60.4       & 64.3 & 54.8  \\
Qwen2.5-1.5B-Instruct~\cite{qwen2.5}    & 61.6      & 49.4          & 63.2 & 55.6     \\
OpenCoder-1.5B-Instruct~\cite{opencoder} & 72.5 & 67.7 & 72.7 & \textbf{61.9} \\
Yi-Coder-1.5B-Chat~\cite{yicoder} & 67.7 & 63.4 & 68.0 & 59.0 \\
Phi-3-Mini-4K-3.8B-Instruct~\cite{phi3technicalreporthighly} & 64.6 & 59.1 & 65.9 & 54.2 \\
CodeGemma-7B-Instruct~\cite{codegemmateam2024} & 60.4 & 51.8 & 70.4 & 56.9 \\
\midrule

\textbf{GrammarCoder-1.3B-Instruct}  & 70.7      & 64.0       & 71.2 & 58.7  \\

\textbf{GrammarCoder-1.5B-Instruct}  & \textbf{73.2}          & \textbf{68.3}           &  \textbf{73.3}    &  61.1     \\ 

\bottomrule
\end{tabular}
}
  \caption{\label{generation}
Performance of various base models and instruct models on HumanEval and MBPP.    
  }
\vspace{-0.5cm}
\end{table*}

\subsection{Experiment I: Performance on Code Generation}\label{ex1}
\paragraph{Evaluation Benchmarks.}
We use HumanEval~\cite{humaneval} and MBPP~\cite{mbpp}, the most widely used datasets for code generation tasks, as evaluation benchmarks. 
HumanEval contains 164 tasks, while MBPP includes 378 testing tasks, both equipped with built-in test cases for evaluation. EvalPlus~\cite{evalplus} extends these datasets by introducing stricter test cases to improve assessment robustness. We conduct evaluations using both the original benchmark test cases and their EvalPlus-enhanced versions, denoted by a ``+'' suffix.

\paragraph{Baselines.}
\lqy{
To evaluate the effectiveness of grammar-based code representation,we select representative and competitive baseline models for comparison, including DeepSeek-Coder-1.3B~\cite{guo2024deepseekcoder}, Qwen2.5-1.5B, and Qwen2.5-7B~\cite{qwen2.5}.
DeepSeek-Coder-1.3B, trained on high-quality large-scale code datasets, serves as a strong representative of the code model. Meanwhile, Qwen2.5-1.5B-Base and Qwen2.5-7B-Base, despite being a general-purpose model, demonstrate competitive performance on code-related tasks, making them a valuable point of comparison.
}

\paragraph{Metrics.}
We adopt Pass@1 as the evaluation metric. Specifically, for each problem, the model generates a single code sample, which is deemed correct if it passes all predefined unit tests. The Pass@1 score is calculated as:
\[
\text{Pass@1} = \frac{\text{Number of problems solved correctly}}{\text{Total number of problems}}
\]

\paragraph{Implementation Details.}
GrammarCoder models are trained with 8 NVIDIA H800 GPUs.
During the continued pre-training phase, we adopt a two-stage learning rate strategy, following approaches from OpenCoder~\cite{opencoder} and MiniCPM~\cite{hu2024minicpm}. Initially, we use a higher learning rate of 3e-4 to accelerate convergence to a reasonable parameter range. The learning rate is reduced to 5e-5 in the annealing stage for further performance optimization. 
During the instruction tuning phase, we set the learning rate to 5e-5 and trained on an instruction dataset to improve generalization in instruction understanding and code tasks. 
Throughout the training process, we apply 100 warm-up steps and use a cosine learning rate scheduler to ensure smooth learning rate adjustments, maintaining training stability and efficiency. Additionally, during both token-based and grammar-based continued pre-training, we utilize the same settings to ensure a fair comparison. 
\ref{training_dataset} and \ref{dataset_filter} provide the detailed information of training dataset.

\paragraph{Results.}
Table~\ref{mainexps} presents our main experimental results, showing that the GrammarCoder-Base model significantly outperforms both the original model and the token-based model trained on the same datasets. For example, on the HumanEval dataset, GrammarCoder-1.3B-Base achieves 82\% and 44\% improvements over DeepSeek-Coder-1.3B-Base and DeepSeek-Coder-1.3B-Base~(CPT), respectively. Similarly, on the MBPP dataset, GrammarCoder-7B-Base outperforms DeepSeek-Coder-7B-Base and DeepSeek-Coder-7B-Base~(CPT) by 10.3 and 3.7 percentage points, respectively.
\lqy{
Notably, after performing continued pre-training on the training dataset, both the token-based and grammar-based models exhibit performance gains. Moreover, even without grammar-based representation, neither the original token-based model nor the continued pre-trained model produces syntax errors, with syntax correctness nearly reaching 100\%. Occasional syntax errors~(fewer than three) only occur due to random variations on the HumanEval and MBPP datasets. 
Despite this near-perfect syntax correctness, the grammar-based model still demonstrates superior performance, indicating that incorporating grammar rules provides additional benefits beyond merely preventing syntax errors.
}

Building on the base model, we further compare our approach with the best-performing models to date and perform instruction tuning to enhance its ability to follow instructions.
Table~\ref{generation} compares the performance of GrammarCoder with current state-of-the-art code models of similar or larger scales. 
Due to limited computational resources, we perform instruction tuning only on the smaller variants~(i.e., GrammarCoder-1.3B-Base and GrammarCoder-1.5B-Base).
Experimental results show that grammar-based code representation achieves performance comparable to the best token-based models. 
For example, on the HumanEval~(+) dataset, both the base and instruct versions of GrammarCoder-1.5B outperform other models~(e.g., CodeGemma-7B and Yi-Coder-1.5B), while the instruct version achieves performance on par with OpenCoder-1.5B-Instruct.
Similarly, on the MBPP~(+) dataset, GrammarCoder-7B-Base outperforms the OpenCoder-8B-Base model, further demonstrating the effectiveness of grammar-based representations at larger model scales.
\lqy{
However, on the MBPP+ dataset, GrammarCoder-Base-1.5B does not surpass OpenCoder-Base-1.5B, which may be attributed to differences in training data volume and quality during the base model pre-training stage. OpenCoder benefits from training on over 900B tokens of high-quality data, whereas GrammarCoder is pre-trained on only around 10B tokens in grammar-based representation. 
This suggests that while grammar-based representation proves to be effective, the scale and quality of training data also play a crucial role in achieving state-of-the-art performance. 
Future work can explore expanding the amount of high-quality code data processed into grammar-based representations to further enhance model performance.
}


\begin{figure*}[t]
\centering
  \includegraphics[scale=0.7]{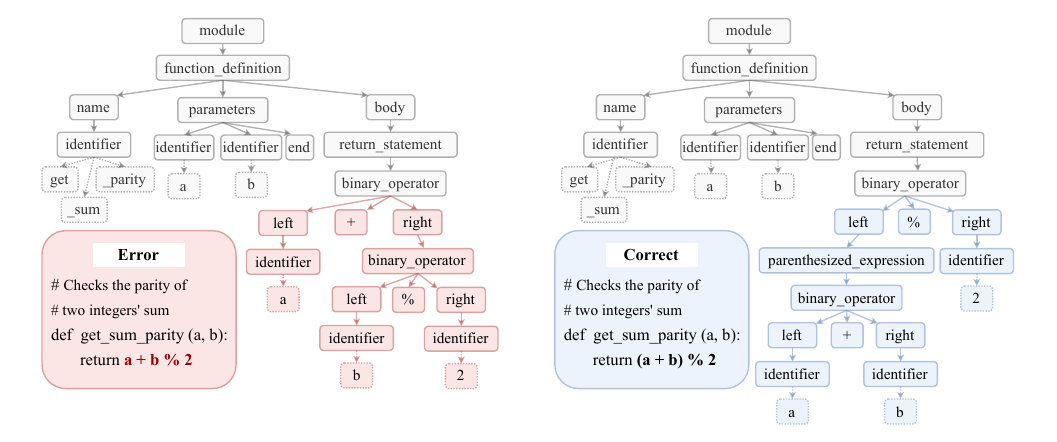}
  \caption {An example showing the differences of code representations between error and correct code.}
  \label{edit_ex}
  \vspace{-0.3cm}
\end{figure*}

\subsection{Experiment II: Understanding the Performance Difference}\label{ex2}



\paragraph{Experiment Design.}

\lqy{
While our experimental results demonstrate that grammar-based representation enhances code generation, it remains crucial to understand what drives this improvement, especially given that syntax errors are already rare in billion-scale LLMs.
To explore the reason behind these results, we focus on why grammar-based representations help mitigate semantic errors beyond preventing syntactic errors, aiming to uncover their role in reducing overall mistakes.
Analyzing the different representation results, we observe that grammar-based representation may amplify differences between correct and incorrect programs that appear minimal at the token level.
This heightened sensitivity to fine-grained variations may help prevent LLMs from behaving like ``careless programmers'', who often make mistakes by overlooking subtle details. By capturing these distinctions more effectively, grammar-based models could reduce such semantic errors, leading to higher performance in code generation tasks.
}

\lqy{
To validate this hypothesis, we design a new set of experiments focusing on subtle semantic changes that are likely to be overlooked by both humans and token-based models. 
Specifically, we investigate (1) whether grammar-based representation amplifies these differences, and (2) whether grammar-based models can better capture these changes.
These experiments aim to provide deeper insight into how grammar-based representation improves the model’s ability to distinguish between correct and error code, making it more effective in avoiding semantic errors and improving performance. 
As we previously observed similar performance trends across different scales of the base models~(i.e., results in Table~\ref{mainexps}), we conduct experiments only at the 1.3B and 1.5B scales in this study to conserve computational resources.
}


First, we conduct a quantitative analysis to explore the potential differences between grammar-based and token-based representations. Specifically, we encode code snippets that are similar at the token level but semantically different using both representation strategies and compare their edit distance when transforming one code into another.  

\lqy{
Next, we train separate grammar-based and token-based code semantic classifiers to evaluate the impact of grammar-based representations on semantic classification. 
By training classification models on differently represented code datasets, we examine the extent to which each representation affects the model's ability to capture semantic differences. 
}

Finally, 
we assess whether the differences introduced by grammar rules contribute to performance improvements, confirming their effectiveness in enhancing LLMs.

\begin{table}
  \centering
  \scalebox{0.7}{
\begin{tabular}{lccc}
\toprule
                            & Precision                 & Recall                    & F1-Score                  \\ \midrule
DeepSeek-Coder-1.3B-Base     & 71.99                     & 62.77                     & 67.06                     \\
DeepSeek-Coder-1.3B-Instruct &  74.20                         &     65.59                      &  69.63                         \\
Qwen2.5-1.5B-Base~\cite{qwen2.5}           &   72.16                        &   64.97                        &  68.38                         \\
Qwen2.5-1.5B-Instruct       &   71.42                        & 67.32                        &  69.31                         \\
Condor-1.3B~\cite{liang2024condor}                 & 74.39 & 72.40 & 73.38 \\ \midrule
\textbf{GrammarCoder-1.3B-Base }     &  \textbf{77.39}                         &       \textbf{81.30 }                   &             \textbf{79.30}              \\
\textbf{GrammarCoder-1.5B-Base}      &  72.34                         &    76.50                       &   74.36                        \\ \bottomrule
\end{tabular}
}
  \caption{\label{understanding}
    The performance of semantic classification tasks.
  }
  \vspace{-0.5cm}
\end{table}

\paragraph{Result 1: Grammar-Based Representation Amplifies Subtle Token-Level Differences.}
We analyze whether grammar-based representation amplifies subtle differences by comparing the edit distances between semantically different code snippets under grammar-based and token-based representations.
CodeNanoFix~\cite{liang2024condor} dataset is used to measure the edit distance, providing a quantitative assessment of how grammar-based and token-based approaches represent code. 
This dataset consists of 1,000+ programming problems and nearly 100,000 code sample pairs with minimal token differences but significant semantic variations. A subset of 120 programming problems and 3,583 sample pairs serve as the test set. 
Each sample in the dataset consists of error code submitted by human programmers while solving a problem, along with its corrected version modified by the programmer, both exhibiting minimal token-level differences. 
Since the differences between error and corrected code typically involve subtle yet crucial semantic changes, such as control flow modifications, variable scope adjustments, and operator usage corrections, this dataset is well-suited for analyzing the differences between token-based and grammar-based representations. 
To mitigate the impact of outliers, we focus on code pairs with minimal token-level differences~(edit distance less than 50, covering 91.18\% of the CodeNanoFix's test set). Additionally, we use GrammarCoder-1.3B's vocabulary to produce grammar-based representations and DeepSeek-Coder-1.3B's vocabulary to produce token-based representations, ensuring a fair comparison is made with the maximum overlap of shared tokens.

The results show that grammar-based representation typically produces larger edit distances compared to token-based representation. 
Specifically, the average edit distance from error to correct code at the token level is 14.33, while for grammar-based representations, it increases to 27.43, a 91.18\% increase. 
\ref{ex_edit_dis} shows the edit distance distribution for error-correct code pairs.
Figure~\ref{edit_ex} presents a concrete example, where the left side shows an error code snippet caused by neglecting operator precedence, while the right side displays the correct version. At the token level, the difference between the two codes consists of only two characters~(i.e., `(' and `)'), resulting in an edit distance of 2. However, at the grammar representation level, the change in operator precedence leads to significant differences in the AST structure and the grammar rules applied, increasing the edit distance to 6.
\ref{ex_model_outputs} also presents the further analysis results of the LLM's generated outputs on CodeNanoFix, which reveal similar conclusions.
These results indicate that the introduction of grammar rules amplifies representation differences that may be overlooked at the token level. Consequently, the grammar-based representation provides a more distinct encoding of correct and incorrect code, allowing the model to better capture semantic variations.


\paragraph{Result 2:  Grammar-Based Representation Strengthens Semantic Distinction.}
We evaluate whether grammar-based models more effectively capture these changes by training classifiers using different code representation approaches.
Specifically, we use CodeNanoFix as a dataset for a semantic classification task, evaluating the model’s ability to distinguish between semantically correct and incorrect code. By training classifiers to identify code correctness, we examine whether grammar-based representations improve the model’s understanding of code semantics.
To ensure a fair comparison, we select baselines that align with GrammarCoder's architecture. Specifically, we use its corresponding base models, DeepSeek-Coder-1.3B and Qwen2.5-1.5B, along with Condor-1.3B~\cite{liang2024condor}, a model specifically designed for the CodeNanoFix classification task.
Precision, Recall, and F1 score are utilized as key metrics for classification performance. Precision evaluates the accuracy of correct code predictions, Recall measures the model’s ability to identify the actual correct code, and F1 score provides a balanced assessment of overall classification performance.
Similar to Condor, GrammarCoder is fine-tuned on the CodeNanoFix dataset to enhance its understanding of code semantics and alignment with problem descriptions. In the implementation, a classification layer is added to the original model to output probability scores, with 0.5 sets as the classification threshold. Code snippets with scores above 0.5 are considered correct, while those below are classified as errors. During fine-tuning, the learning rate is set to 5e-5 to ensure stable optimization for the code classification task.

Table~\ref{understanding} illustrates the impact of different code representation approaches on the model's ability to determine code semantic correctness.
The results indicate that incorporating grammar rules significantly enhances the model’s ability to distinguish correct from incorrect code. For example, GrammarCoder-1.3B-Base and GrammarCoder-1.5B-Base improve F1 scores by 18.25\% and 8.75\%, respectively, compared to their base models DeepSeek-Coder-1.3B-Base and Qwen-1.5B-Base. 
These results demonstrate that the incorporation of grammar rules enables the model to more precisely differentiate token-level similar but semantically distinct code snippets, improving its ability to recognize subtle semantic differences.
Furthermore, even compared to Condor, the current best-performing model on the CodeNanoFix dataset, GrammarCoder-1.3B-Base achieves nearly nine percentage points higher Recall and improves the F1 score by almost six percentage points.
Notably, both Condor and GrammarCoder-1.3B-Base are trained from the same baseline model, DeepSeek-Coder-1.3B-Base.
This further highlights the effectiveness of grammar-based representation in distinguishing semantic differences caused by subtle token-level changes in code.

\paragraph{Result 3: Correlation Between Representation and Performance.}
We conduct a correlation analysis to examine whether the increase in edit distance is related to GrammarCoder's ability to distinguish between semantically correct and incorrect code.
A chi-square test confirms a statistically significant correlation, with GrammarCoder-1.3B-Base and GrammarCoder-1.5B-Base achieving p-values of 0.0051 and 0.0006, respectively. As a p-value below 0.05 indicates statistical significance, the results suggest that grammar-based representation contributes to performance improvements by amplifying structural differences in code. 
\ref{generation_cases} also presents case studies where the token-based model’s generated outputs can be corrected with minor modifications at the token level.
This ability brought by grammar-based representation helps prevent the model from exhibiting oversight-prone tendencies akin to a ``careless programmer,'' where minor but critical details are ignored, potentially leading to semantic errors.
As a result, grammar-based representation not only improves the model's understanding of code semantics but also enhances overall performance in code generation.

\section{Related Work}
\subsection{Large Language Models for Code}
Since the release of ChatGPT-3.5~\cite{chatgpt3.5} sparked a new wave of interest in LLMs, increasing focus has been on training and utilizing LLMs for code-related tasks~\cite{chen2025sesurvey, wang2024prompt}. These models can be broadly categorized into two types. 
The first category consists of general-purpose models, which perform well in various natural language tasks, while also showing strong capabilities in code-related tasks. 
Examples of models in this category include ChatGPT~\cite{chatgpt2024}, Gemini~\cite{gemini}, Claude~\cite{claude}, Qwen~\cite{qwen2.5}, and DeepSeek~\cite{bi2024deepseek}. 
The second category comprises models specifically trained on code data, including models such as CodeLlama~\cite{codellama}, OpenCoder~\cite{opencoder}, and DeepSeek-Coder~\cite{guo2024deepseekcoder}. Compared to general-purpose models, these specialized models can achieve comparable or superior performance on code-related tasks with fewer parameters and offer broader support for less common programming languages.
However, regardless of whether they have been specifically trained for code-related tasks, these models represent programming languages in the same way as natural language—using token sequences. 
This hinders the model's ability to recognize the inherent structural information of programming languages. 
Therefore, we leverage the grammar-based code representation to train GrammarCoder, which enhances the model's ability to capture structural information inherent in programming languages.


\subsection{Grammar-Based Code Representation}


Many models attempt to incorporate grammar-based information into code representations~\cite{sun2019grammar, liangbipartite, unixcoder, grammart5, treegen, l2s, abstract}. These models have been validated on relatively small-scale models~(fewer than 220M parameters), demonstrating that grammar-based representation helps prevent syntax errors and enhances code generation performance.
For example, GrammarT5~\cite{grammart5} is a pre-trained model based on grammatical rules. It is trained based on CodeT5~(220M)~\cite{codet52021} with an encoder-decoder architecture using the same training data, demonstrating that grammar-based representations can enhance model performance.
However, with the emergence of LLMs, models' size has expanded rapidly, and decoder-only architectures have gradually become mainstream. It's unclear whether grammar-based representations remain effective in larger-scale~(e.g., billion-size)~decoder-only models. Moreover, beyond preventing grammatical errors, it remains unclear whether grammar-based representations provide any additional benefits. 
Therefore, we bridge these gaps by training and evaluating grammar-based representations in billion-scale decoder-only models. Additionally, we explore why grammar-based representation remains effective when syntax errors are rare in LLMs, providing insights into its broader impact on model performance.

\section{Conclusion}
In this paper, we introduce GrammarCoder, a series of models trained using grammar-based code representations. 
To evaluate whether this approach remains effective even when billion-scale models basically no longer make syntax errors, we assess GrammarCoder on widely used code generation benchmarks, HumanEval(+) and MBPP(+). Experimental results show that after continued pre-training on the same datasets, GrammarCoder significantly outperforms models trained with normal token-based representations.
To further investigate why grammar-based code representations are effective, we first quantify the differences between grammar-based and token-based approaches in representing code. Additionally, we train a classification model to assess their ability to capture subtle code variations. Our findings reveal that while modern LLMs rarely make syntax errors, grammar-based representations still enhance their ability to distinguish fine-grained token-level differences. This reduces semantic errors caused by minor variations and ultimately improves model performance in code-related tasks.

\section{Acknowledgment}
This work is sponsored by the National Key Research and Development Program of China under Grant No. 2023YFB4503803, and the National Natural Science Foundation of China under Grant No. W2411051, No. 62232003, and No. 62402482. Kuaishou Technology also supports this work.

\section{Limitations}
While grammar-based representations excel in code understanding and generation, they might face the following limitations.  
First, they may struggle with non-standard or incomplete code. Real-world datasets often contain code mixed with natural language or truncated snippets, which may fail AST parsing, reducing data utilization.  
Second, grammar-based models may struggle with incomplete syntax. 
When dealing with incomplete variable names or missing key symbols~(e.g., brackets, commas), grammar-based approaches may face higher parsing or pre-processing costs.
In these cases, token-based approaches offer greater flexibility.

Generally, grammar-based representation remains effective in billion-scale LLMs, enhancing the model’s ability to capture subtle semantic changes. This leads to improvements in code generation and semantic classification accuracy. However, its reliance on AST parsing introduces challenges in processing incomplete or syntactically incorrect code, limiting its flexibility.



\bibliography{custom}

\appendix

\section{Approach Details}
\subsection{Mode Configuration}\label{mode_config}

\begin{table}[h]
\centering
\scalebox{0.7}{
\begin{tabular}{lccc}
\toprule
    \textbf{Config}                  & \textbf{DeepSeek-Coder} & \textbf{Qwen2.5} &\textbf{Qwen2.5}  \\ \midrule 
\# parameters         & 1.3 B       & 1.5 B & 7 B \\
\# hidden\_layer    & 24        & 28    & 28       \\
\# hidden\_size       & 2,048       & 1,537 & 3,584        \\
\# intermediate\_size & 5,504   & 8,960 & 18,944       \\
\# attention\_head   & 16          & 12 &  28        \\
\# vocabulary         & 32,256      & 151,936 &  152,064    \\ \bottomrule
\end{tabular}
}
\caption{\label{basemodel}
The main configuration of base models.
}
\end{table}

Table~\ref{basemodel} presents the key configurations of the base models used in our study: DeepSeek-Coder, Qwen2.5-1.5B, and Qwen2.5-7B. While these models are billion-scale in terms of parameter count, they exhibit differences in architectural details, particularly in vocabulary size. DeepSeek-Coder uses a vocabulary of 32,256 tokens, whereas Qwen2.5 adopts a significantly larger vocabulary of over 150,000 tokens.
Since grammar-based representations restructure code at a syntactic level rather than relying solely on the token level, their effectiveness is not dependent on the original vocabulary. 
Therefore, this difference in vocabulary size can underscore the robustness of our grammar-based code representation. 
After incorporating grammar rules, our vocabulary sizes expand to 33,465 for DeepSeek-Coder, 153,108 for Qwen2.5-1.5B, and 154680 for Qwen2.5-7B.
If GrammarCoder demonstrates improved performance across both base models, it would further indicate that grammar-based approaches are adaptable to different model architectures and tokenization strategies.

\subsection{Training Objective}\label{ex_training_objective}
The training objective of GrammarCoder is to maximize the conditional probability of the next token given the preceding sequence. The loss function of training objective can be formalized as:

\[
\mathcal{L} = - \sum_{t=1}^{N} \log P(x_t | x_1, x_2, \dots, x_{t-1}; \theta)
\]
, where \( x_t \) represents the token (either a normal token or a grammar rule from \( V_{\text{grammar}} \)) at step \( t \), \( x_1, x_2, ..., x_{t-1} \) denotes the previously generated sequence, \( \theta \) represents the model parameters, and the objective is to maximize the conditional probability of the correct token given the current context \(P(x_t \mid x_1, x_2, ..., x_{t-1})\).

This ensures that the final output adheres to syntax constraints while effectively capturing correct program logic, aligning with the preorder traversal of the complete AST.

\subsection{Training Datasets and Filter}\label{dataset_filter}

\begin{table}[h]
\centering
\scalebox{0.8}{
\begin{tabular}{lc}
\hline
\textbf{Name}                                    & \# \textbf{Samples} \\ \hline
code\_contests\_instruct                & 4.4 M      \\
Opencoder-sft-stage1                    & 4.2 M      \\
Opencoder-sft-stage2                    & 375K       \\
Code-290k-ShareGPT-Vicuna-Clean         & 285K       \\
CodeFeedback-Filtered-Instruction       & 156K       \\
code\_instructions\_122k\_alpaca\_style & 121K       \\ \hline
\end{tabular}
}
\caption{\label{dataset}
Open-source instruction datasets.
}
\end{table}

\begin{figure}[t]
    \centering
  \includegraphics[width=0.75\linewidth]{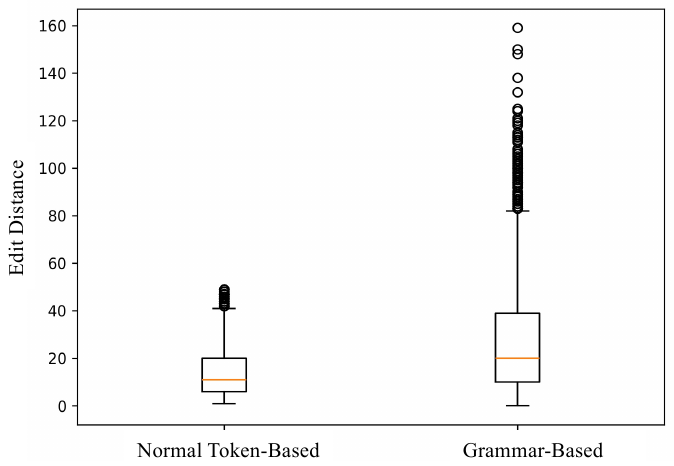}
  \caption {Edit distance distribution across different code representation approaches.}
  \label{edit}
\end{figure}

\begin{figure*}[t]
  \includegraphics[width=0.45\linewidth]{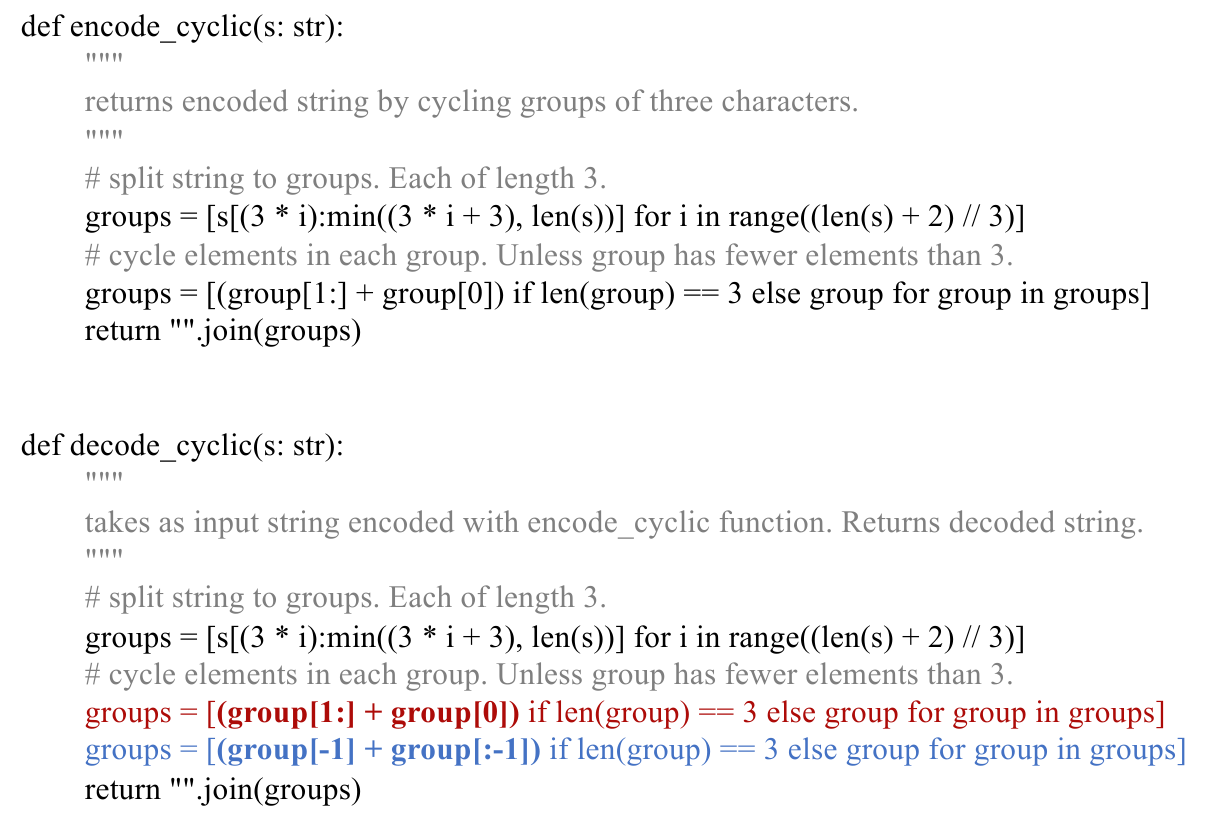} \hfill
  \includegraphics[width=0.55\linewidth]{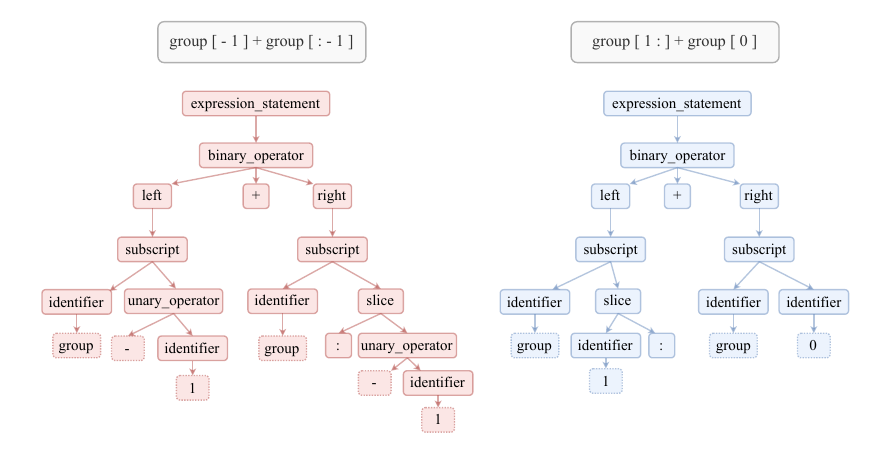}
  \caption {
  DeepSeek-Coder-1.3B-Base~(CPT)'s generated output for Task 38 in the HumanEval dataset (left) and the required AST modifications to correct the code (right).}
  \label{he38}
\end{figure*}

\begin{figure*}[t]
  \includegraphics[width=0.45\linewidth]{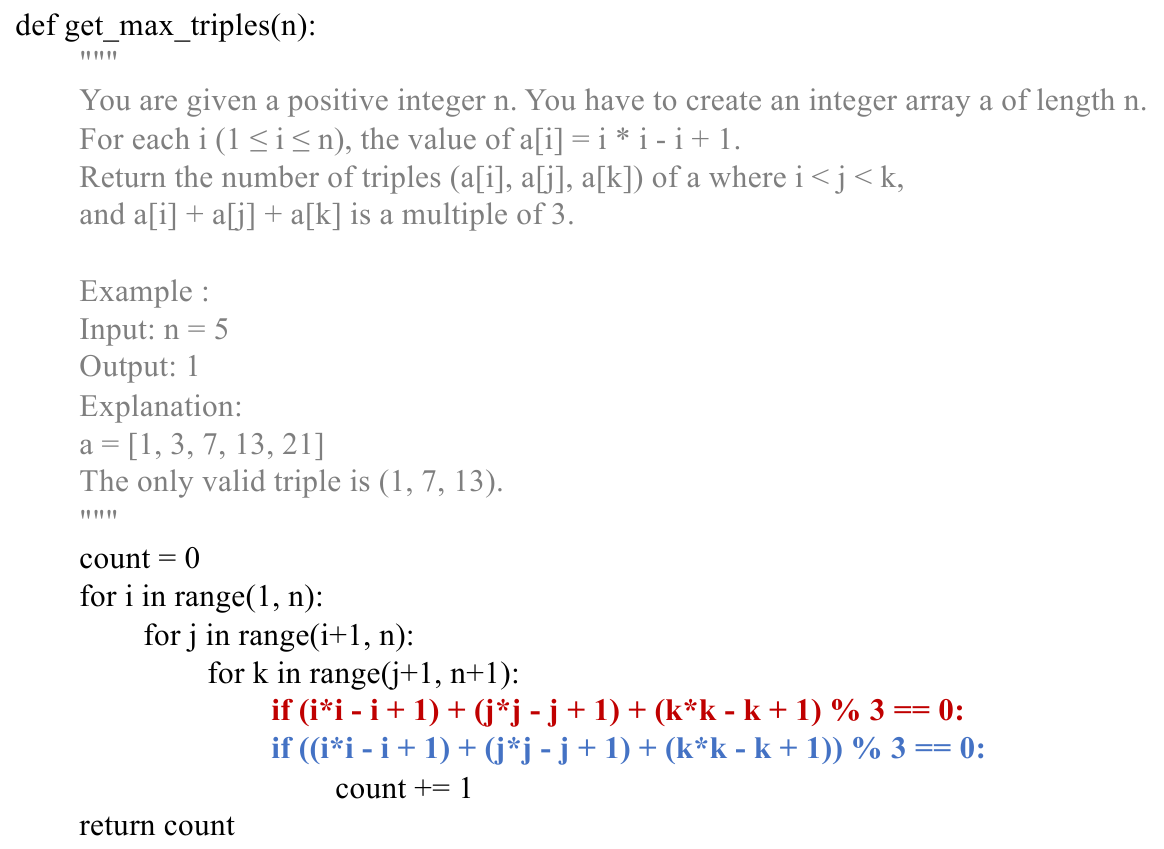} \hfill
  \includegraphics[width=0.55\linewidth]{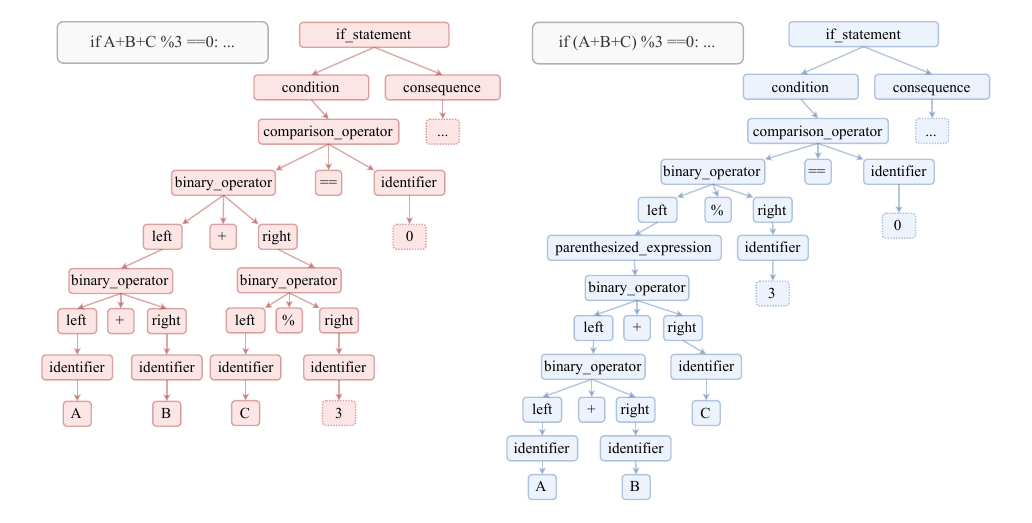}
  \caption {DeepSeek-Coder-1.3B-Base~(CPT)'s generated output for Task 147 in the HumanEval dataset (left) and the required AST modifications to correct the code (right). For clarity, we represent identical computational units before and after modification using A, B, and C, respectively.}
  \label{he147}
\end{figure*}

For training the base models, we primarily use high-quality Python code, aligning with our focus on grammar-based code representation. Our dataset is composed of two key sources. First, we sample 10B tokens from TheStackV2~\cite{starcoder2}, a large-scale code corpus that provides diverse and high-quality programming samples across various domains, ensuring a strong foundation in general coding patterns and structures. Second, inspired by previous studies~\cite{opencoder,qwen2}, we incorporate 0.5B tokens of self-contained code textbooks from open-source repositories~\cite{opencoder}. Unlike context-dependent snippets, these samples consist of independent tasks and corresponding code snippets, helping the model learn to generate independent and coherent programs, bridging the gap between standard pre-training and real-world interactive programming scenarios.

For training the instruct models, we use instruction data consisting of two main sources: publicly available instruction datasets and synthetically generated instruction data. Table~\ref{dataset} lists the open-source instruct datasets used in our training~\cite{hu2024minicpm, opencoder, codesharegpt, zheng2024codefeedback, codealpaca}, each contributing to the diversity and quality of instruction tuning.
All of the datasets have a permissive license for the training LLM.
For synthetic instruct data, we use LLaMA3.1-70B as the base model to generate high-quality data, leveraging OSS-Instruct~\cite{wei2024magicoder} and SelfCodeAlign~\cite{wei2024selfcodealign} as synthesis methods. This approach enables us to create a large-scale instruct dataset totaling 5B tokens, further enhancing the model’s ability to follow instructions effectively.

To ensure data quality, we apply data filtering for both base models and instruct models, primarily focusing on deduplication and syntax validation. Deduplication is performed through string-based text matching to eliminate redundant samples. For syntax validation, we use Tree-sitter~\cite{treesitter} to check whether the code can be parsed into a valid syntax tree; if parsing fails, the sample is removed. These filtering steps help maintain a high-quality dataset for training.



\section{Experimental Details}
\subsection{Distribution of Edit Distance}\label{ex_edit_dis}

Figure~\ref{edit} shows the edit distance distribution for error-correct code pairs with small edit distances~(less than 50, accounting for 91.18\% of the test set) under different representation approaches.

\subsection{Analysis of Model Outputs}\label{ex_model_outputs}
While we have examined the differences between representations on existing datasets, it is also crucial to analyze whether grammar-based representation amplifies token-level subtle differences in the model’s generated outputs. 
Therefore, we further analyzed the inference results of Meta-Llama-3.1-70B~\cite{llama3} on the CodeNanoFix dataset, focusing on the edit distance between correct and incorrect code samples for the same data samples. 
The results show that in 25.56\% of the samples, the token-level edit distance between incorrect and correct code is relatively small~(less than 50). Among these samples, the average edit distance for token-based representations is 28.04, whereas for grammar-based representations, it increases to 44.56.  
These findings suggest that even for a 70B-scale model, generating the correct code remains challenging when token-level differences are minimal. Relying solely on token-level information may not be sufficient to distinguish critical semantic differences in code. In contrast, grammar-based representations provide additional structural information, helping the model better differentiate between similar yet semantically distinct code snippets.

\subsection{Errors caused by subtle differences.}\label{generation_cases}

Figures~\ref{he38} and \ref{he147} illustrate errors made by the token-based LLM~(DeepSeek-Coder-1.3B-Base (CPT)) on the HumanEval dataset, highlighting how these mistakes can be corrected with minimal token-level modifications. For example, in Figure~\ref{he38}, fixing the error requires only adjusting the range of operations within the `group' list, while in Figure~\ref{he147}, the bug can be fixed by adding a single pair of parentheses to enforce the correct order of operations.

However, since these examples require only minor token-level modifications, they may be overlooked by token-based LLMs. In contrast, grammar-based representations introduce larger structural changes in the corresponding AST, making the model more sensitive to differences between correct and incorrect code. 
These examples demonstrate that grammar-based models, by explicitly organizing code through grammar rules, can better capture subtle code variations. As a result, grammar-based models are more effective in recognizing and generating correct code, even in cases where small token-level changes drastically alter program behavior.

\end{document}